\documentclass[fleqn,10pt]{wlscirep}
\title{Determination of Asphaltene Critical Nanoaggregate Concentration Region Using Ultrasound Velocity Measurements}

\author[1,*]{Aleksandra Svalova}
\author[2]{Nicholas G Parker}
\author[3]{Malcolm JW Povey}
\author[1]{Geoffrey D Abbott}
\affil[1]{Newcastle University, School of Natural and Environmental Sciences, Newcastle upon Tyne, NE1 7RU, United Kingdom}
\affil[2]{Newcastle University, School of Mathematics, Statistics and Physics, Newcastle upon Tyne, NE1 7RU, United Kingdom}
\affil[3]{University of Leeds, School of Food Science and Nutrition, Leeds, LS2 9JT, Leeds, United Kingdom}
\affil[*]{a.svalova@newcastle.ac.uk}
\begin{abstract}
Asphaltenes constitute the heaviest, most polar and aromatic fraction of petroleum crucial to the formation of highly-stable water-in-crude oil emulsions. The latter occur during crude oil production as well as spills and cause difficulties to efficient remediation practice. It is thought that in nanoaggregate form, asphaltenes create elastic layers around water droplets enhancing stability of the emulsion matrix. Ultrasonic characterisation is a high-resolution non-invasive tool in colloidal analysis shown to successfully identify asphaltene nanoaggregation in toluene. The high sensitivity of acoustic velocity to molecular rearrangements and ease in implementation renders it an attractive method to study asphaltene phase properties. Currently, aggregation is thought to correspond to an intersection of two concentration-ultrasonic velocity regressions. Our measurements indicate a variation in the proximity of nanoaggregation which is not accounted for by present models. We attribute this uncertainty to physico-chemical heterogeneity of the asphaltene fraction driven by variation in molecular size and propose a critical nanoaggregation region. We treated asphaltenes from North and South American crude oils with ruthenium ion catalysed oxidation to characterise their \textit{n}-alkyl appendages attached to aromatic cores. Principal component analysis was performed to investigate the coupling between asphaltene structures and velocity measurements and their impact on aggregation.
\end{abstract}
\begin{document}

\flushbottom
\maketitle
\thispagestyle{empty}

\section*{Introduction}
Water-in-oil emulsions (WOE) are highly stable mixtures occurring during crude oil production and spills~\cite{bridie1980,sjoblom2003,dicharry2006}. Emulsion removal from the water column is problematic due to the inherent high elasticity, viscosity and stability of these mixtures~\cite{sjoblom2003,dicharry2006}. For efficient removal, WOEs require separation into water and oil phases~\cite{dicharry2006,singh1999} which is inhibited by the presence of asphaltenes at the water/oil boundary~\cite{sjoblom2003,dicharry2006,singh1999,mullins2011}.\\
Asphaltenes are the heaviest, most aromatic and polar constituents of crude oil, soluble in aromatic solvents and precipitating upon the addition of low molecular weight {\it n}-alkanes \cite{sheu1995,mullins2007,sjoblom2003,sjoblom2015}. Their solubilisation may change depending on solvent~\cite{peters2005c} and the number of aromatic rings~\cite{friberg2007}. The asphaltene molecules have a wide distribution of architectures with two broad types, island and archipelago~\cite{mullins2010}, recently thought of as two extremes of a structural continuum~\cite{durand2010,dasilvaoliviera2014,hosseini2015}. The island architecture is most common ~\cite{mullins2010,schuler2015,schuler2017} and comprises a single polycyclic aromatic hydrocarbon (PAH) centre and aliphatic appendages~\cite{mullins2010,schuler2015}. The archipelago architecture is characterised by several (smaller) PAHs connected by an aryl linkage~\cite{schuler2017}. Recent studies using atomic force microscopy~\cite{schuler2015,schuler2017} presented images of asphaltene coal and petroleum specimens and confirmed their PAH-side chain architecture as well as the high structural variability; less than 10\% of specimens were archipelago. Studies on model compounds and asphaltenes using two-step laser mass spectroscopy~\cite{sabbah2011,mullins2011} and laser-induced acoustic desorption electron impact mass spectroscopy~\cite{borton2010,mullins2011} concluded that island molecules remained stable under different conditions whereas archipelago structures were susceptible to fragmentation, which may be the reason for their low abundance. A typical mass of an island monomer is 750 Da ($\pm$ 250 Da) and archipelago specimens may reach 2000 Da~\cite{mullins2011,groenzin1999,groenzin2000,buenrosto2001,groenzin2003,buch2003}. The Yen-Mullins model is most widely used in describing asphaltene aggregation and structure~\cite{mullins2010,mullins2011}. At concentrations below ca. 50 mg/L, asphaltenes are believed to exist as single molecules or `monomers'. As concentrations increase to the critical nanoaggregate concentration (CNAC) of ca. 100 mg/L ($\pm$ 50 mg) asphaltenes start forming nanoaggregates. In this form (or at concentrations above CNAC) asphaltenes are believed to stabilise WOEs by creating a `skin' around water droplets~\cite{singh1999,yarranton2000,jestin2007,mullins2011,friberg2007a,fan2010}. At concentrations of 2-5 g/L nanoaggregates begin forming clusters, each consisting of less than ten nanoaggregates~\cite{goual2011,mullins2011}. In our investigation the concentrations at which asphaltenes are studied are below cluster formation. Previous works regarding asphaltene aggregation include the Yen model~\cite{yen1961,dickie1967}, fractal model~\cite{barre2008,simon2009} and colloidal model~\cite{pfeiffer1940}.\\
In what follows we investigate the aggregation properties of four asphaltene samples using ultrasonic velocity precision measurements of asphaltene-in-toluene solutions, ruthenium ion catalysed oxidation (RICO) and statistical analysis. We also perform a biodegradation characterisation of asphaltene parent oils. Our velocity characterisation results suggest that asphaltene aggregation occurs over a critical nanoaggregation region (CNR), rather than at a fixed CNAC. We review the theory of surface-active compound (surfactant) aggregation to highlight phenomena explaining asphaltene phase behaviour. In particular, multiple micellarisation can be useful in understanding aggregation of multi-sized surfactant systems. We suggest that the dimensions of CNR are related to asphaltene side-chain distribution which we obtain using ruthenium ion catalysed oxidation (RICO)~\cite{peng1999,snowdon2016}. This method induces selective oxidation of aromatic rings, releasing the aliphatic chains almost unaltered. In order to discern the CNR boundaries, we use a statistical constrained optimisation method~\cite{bertsekas1982,smith1997} with linear models, the details of which are presented in Supplementary Information. Our results suggest that the asphaltene concentration range has three regions: (i) linear monomeric, (ii) non-linear critical nanoaggregation (CNR) and (iii) linear aggregated. The Discussion section investigates the relationship between combined asphaltene velocity and structural data using principal component analysis (PCA).
\section*{Materials and methods}
\subsection*{Theory of surface-active compound aggregation}
In the context of WOE formation and remediation, asphaltenes are significant due to their surface activity and emulsion stabilisation properties~\cite{sjoblom2003,mclean1997,grutters2007}. Thus, asphaltenes are often likened to surfactants~\cite{andreatta2005} and below we outline a number of important properties to compare with asphaltene phase behaviour. Surfactants tend to be located at the interface between two liquid phases as this corresponds to their lowest energy state~\cite{clint1992a}. Such interfacial activity agrees with the phenomenon of asphaltene `skin’ formation around water droplets~\cite{singh1999,yarranton2000,khristov2000,jestin2007,mullins2011,rane2012,rane2013}. Molecular dynamic simulations of asphaltenes illustrated that model compounds with highest surface activity (with charged functional groups) were located at the toluene/water interface after 7 ns of simulation (mostly in cluster form), whereas non-charged moieties did not show interfacial activity and mostly remained in bulk toluene~\cite{kuznicki2008,kuznicki2009}. The latter study developed compounds that were structurally similar to the atomi force microscopy (AFM) images from~\cite{schuler2015}, thus the results are representative. Simulations of non-charged asphaltene model compounds illustrated that such asphaltenes were a lot more likely to remain in bulk oil rather than travel to the water/oil boundary~\cite{teklebrhan2014}. Nanoaggregation (self-association) of asphaltenes occurs upon reaching the CNAC, whereby crucial to asphaltene solubilisation are steric repulsion of alkane substituents and $\pi-\pi$ stacking of aromatic cores~\cite{buenrosto2001,zhang2003,andreatta2005}. Steric hindrance restricts the number of asphaltenes in a nanoaggregate and a further addition of asphaltenes to the system will lead to a change in aggregate number, as opposed to size~\cite{friberg2007,andreatta2005}. In other words, nanoaggregation kinetics are controlled by asphaltene (poor) solubility~\cite{buckley2007} and strong asphaltene-asphaltene interactions~\cite{friberg2007}; in an aqueous environment micellarisation is controlled by solvent-surfactant interactions. Interfacial tension (IFT) is one of the most widespread measurements in analysing surfactant emulsion-stabilisation properties~\cite{friberg2007}. Obtaining a plot of surface tension versus the logarithm of surfactant concentration will illustrate a decreasing trend and the CMC. Noteworthy, interfacial tension measurements require that the activity coefficient $a_s$ in the Gibbs adsorption equation (estimating interfacial tension) is constant, which is true for many aqueous systems~\cite{friberg2007}. It may be of use to understand the challenges in applying IFT to asphaltene systems. Firstly, asphaltene activity coefficient is reported to have contradictory properties, whereby some studies report an approximation of unity using Scatchard-Hildebrand solubility theory with a Flory-Huggins term~\cite{yarranton1996}, and other suggest that it is not constant~\cite{friberg2007}. The surface tension of toluene is two and a half times lower than that of water, and loading high-energy asphaltenes onto toluene surface may increase surface tension~\cite{mostowfi2009}. On the other hand, in Langmuir-Blodgett~\cite{friberg2007a} and pendant droplet~\cite{rane2012,rane2013} experiments, asphaltenes were shown to decrease surface tension. Studies by Rane, \textit{et al.}~\cite{rane2012,rane2013} illustrated that in water-heptol systems with 10-50 ppm asphaltene fraction dynamic interfacial tension decreased with no reported asymptotic value, whereas emulsions with 50-200 ppm asphaltene illustrated an asymptotic limit at 20 mN/m. Interestingly, the 50-200 ppm asphaltene concentration is consistent with our estimations of the CNR, and their nuclear magnetic resonance (NMR) data illustrates a linear dependence of asphaltene concentration on NMR signal breaking around a similar 80-200 ppm range. Contradicting previous studies~\cite{mullins2011,friberg2007a}, Rane, \textit{et al.}~\cite{rane2012,rane2013,rane2015} used the Langmuir equation of state to suggest that it is asphaltene monomers that stabilise water-in-model oil emulsions rather than nanoaggregates although their NMR analysis suggested that nearly half of molecules present in model oil were nanoaggegates~\cite{rane2012}.
Despite the above differences, using micellarisation as a proxy for asphaltene nanoaggregation allows the use of sonic velocity models~\cite{zielinski1986,andreatta2005} detecting aggregation as a function of changes in solution apparent densities and compressibilities of the dispersed phase. The full model derivation may be found in Zielinski {\it et al.}~\cite{zielinski1986}, and is summarised as follows.\\
Within a uniform liquid, the ultrasonic velocity $u$ is related to density $\rho$ and adiabatic compressibility $\beta$ of the medium according to the Urick equation~\cite{urick1947}
\begin{equation}\label{urick}
u=\sqrt{\frac{1}{\rho\beta}}.
\end{equation} 
For multi-phase fluids which are well-dispersed, and ignoring the effects of sound scattering (valid for sufficiently low concentration of scatterers and away from scattering resonances)~\cite{povey1997}, Equation~\eqref{urick} can be applied with density and compressibility represented by weighted averages of the mixture components. An extension of Equation~\eqref{urick} allows to detect the onset of surfactant aggregation into micelles, as proposed by Zielinski {\it et al.}~\cite{zielinski1986}. In particular, the sound velocity $u$ is related to apparent molar solution quantities following the relation
\begin{equation}\label{ziel}
u=u_0+\frac{u_0}{2}\left(\tilde{v}_1\left(2-\frac{\tilde{\beta}_1}{\beta_0}\right)-v_0\right)c_1+\frac{u_0}{2}\left(\tilde{v}_m\left(2-\frac{\tilde{\beta}_m}{\beta_0}\right)-v_0\right)c_m,
\end{equation}
where $v$ denotes specific volume, $c$- weight concentration, tilde- apparent quantities and subscripts refer to solvent (0), monomer (1) and micellar (m) quantities. Also,
\begin{equation}
\begin{cases}
\text{if}\,\, c\leq \text{CMC},\,\,\text{then}\,\, c_1=c, \,\,\text{and}\,\,c_m=0,\,\, \text{otherwise}\\
\text{if}\,\, c > \text{CMC}, \,\,\text{then}\,\, c_1=\text{CMC},\,\,\text{and} \,\, c_m=c-\text{CMC}.
\end{cases}
\end{equation}
The model~\eqref{ziel} implies that pre- and post-micellarisation, sonic velocity is related to surfactant concentration as a combination of two linear behaviours whose intersection estimates the CMC. Apparent molar properties are difficult to measure in practice and we presume that an accurate determination of the CMC is found by an intersection of two linear regressions, the optimal selected by informing the coefficient of determination $R^2$. Zielinski \textit{et al.}~\cite{zielinski1986} verified this model by measuring the speed of sound in solutions of alkyltrimethylammonium bromides and illustrated a very good fit. A study by Andreatta \textit{et al.}~\cite{andreatta2005} used high-\textit{Q} ultrasonic velocity measurements together with model in~\cite{zielinski1986} to study asphaltene nanoaggregation in toluene, whereby regressions could describe variation in pre- and post-CNAC regions well. Upon closer examination of the monomer-aggregate boundary the point of asphaltene nanoaggregation is associated with velocity fluctuations, the reasons for which we aim to investigate in more detail. \\
Further, we consider the phenomenon of multiple micellarisation whereby a surfactant solution is prone to forming micelles of multiple sizes given the variation in surfactant molecular size~\cite{clint1992b,ray2005,chen1986}. The latter occurs in some pure substances and is very common in mixtures of cationic surfactants with a variation in the hydrophobic tail length ~\cite{ray2005,chen1986,treiner1992}. Critically, as asphaltenes constitute a fraction of petroleum with a distribution of molecular weights and structures we suggest that a phenomenon similar to multiple micellarisation will also be observed. We will attempt to illustrate the latter by performing ultrasonic characterisation of alkyltrimethylammonium bromide surfactants in single and mixed forms followed by asphaltene dispersions in toluene.

\subsection*{Materials}
Chemical solvents toluene, acetonitrile, and \textit{n}-pentane were purchased from Sigma Aldrich and Acros Organics, all $\geq 99\%$ purity. Dichloromethane (DCM), methanol (MeOH) and petroleum ether were distilled in-house. The surfactants used in verification studies were tetradecyltrimethylammonium bromide (CH\textsubscript{3}(CH\textsubscript{2})\textsubscript{13}N(CH\textsubscript{3})\textsubscript{3}Br; C\textsubscript{14}TAB) and dodecyltrimethylammonium bromide (CH\textsubscript{3}(CH\textsubscript{2})\textsubscript{11}N(CH\textsubscript{3})\textsubscript{3}Br; C\textsubscript{12}TAB) 99\% and 98\% purity respectively, obtained from Sigma Aldrich. The two compounds differ by only two methyl groups and their mixtures were used to test sensitivity of the ultrasonic instrument to multiple micellarisation. Milli-Q 18 M$\Omega$ deionised water was used to make C\textsubscript{14}TAB and C\textsubscript{12}TAB solutions.\\
Asphaltenes were precipitated from four petroleum samples: E1 with E2 and E3 with E4 are from two different source rocks respectively and all are from different reservoirs. E1 and E2 are from South America and E3 with E4 are from North America. Samples were selected such that there are two biodegradation aliquots from a source rock. Biodegradation analysis was performed on deasphalted petroleum using thin layer chromatography with short column elution to obtain aliphatic and aromatic fractions. The latter were then analysed by gas chromatograohy-mass spectrometry and relative biomarker abundance compared with Wenger \textit{et al.}~\cite{wenger2002,peters2005b} scales. 
\subsection*{Asphaltene precipitation}
Asphaltene precipitation was performed using a 40-fold excess \textit{n}-alkane addition~\cite{yu2014,groenzin1999}. Crude oil (5 g) was mixed with 200 ml of \textit{n}-pentane, ultrasonicated for 2 h and left to equilibrate overnight. The mixture was then centrifuged for 15 min at 3500 rpm and maltene supernatant decanted. Further, the asphaltene fraction was washed following a cycle of (i) 200 ml \textit{n}-pentane addition, (ii) ultrasonication for 30 min, (iii) equilibration for 1 h, (iv) centrifugation for 15 min at 3500 rpm and (v) decanting of the supernatant. Asphaltenes were then air-dried overnight and/or dried with nitrogen gas before purification using Soxhlet extraction with toluene~\cite{yu2014}. The obtained asphaltene-toluene solution was evaporated under reduced pressure to 2-5 ml and washed for a final time as described above (i-v) for further use in ultrasonic characterisation and oxidation experiments.
\subsection*{Ruthenium ion catalysed oxidation of asphaltenes}
Inference about asphltene architecture was performed using ruthenium ion catalysed oxidation (RICO)~\cite{peng1999}. The reaction products are homologous series of {\it n}-alkanoic fatty acid methyl esters (FAMEs) representing {\it n}-alkyl appendages that were attached to PAHs and $\alpha,\omega$-\textit{di}-$n$-alkanoic fatty acid \textit{bis}-methyl esters (DFAMEs) representing \textit{n}-alkenyl bridges between two aromatic units~\cite{strausz1999,peng1999,snowdon2016}.
Asphaltenes (50 mg) were mixed with 4 ml DCM, 4 ml acetonitrile, 5 ml 12\% aqueous sodium periodate (NaIO$_4$) and 5 mg ruthenium trichloride (RuCl\textsubscript{3}$\cdot$xH\textsubscript{2}O). The mixture was shaken for at least 18 h using an orbital shaker. Dichloromethane and MeOH (15 ml each) were added to the mixture, shaken vigorously and centrifuged for 15 min at 3500 rpm; this cycle was repeated four times. The supernatant fractions obtained from centrifugation were combined in a separating funnel with 5 ml of 4\% aqueous sodium hydroxide (NaOH), shaken vigorously and left to separate for 30 min. The aqueous phase was washed with DCM. The obtained organic fraction was mixed with 5 ml 13.5\% hydrochloric acid (HCl) and washed with DCM further three times to obtain the final organic phase. This was then evaporated under reduced pressure until dryness, mixed with 5 ml 98:2 mixture of MeOH and sulfuric acid (H$_2$SO$_4$) and reflux-heated for 3h. The obtained esters were mixed with 10 ml deionised water and washed with DCM four times. Finally, the washings were mixed with 4 ml 2\% aqueous NaHCO\textsubscript{3} and evaporated under reduced pressure with sodium sulphate (Na\textsubscript{2}SO\textsubscript{4}) to ca. 5 ml. The products were pipetted out avoiding aqueous traces and blown down with N\textsubscript{2} gas to 1 ml for gas-chromatography-flame ionisation detection (GC-FID) and gas chromatography-mass spectrometry (GC-MS).
\subsection*{Maltene analysis}
Maltene (deasphalted) petroleum samples were separated into aliphatic/saturate, aromatic and polar fractions using thin layer chromatography (TLC) and short column elution. Glass plates were covered with 0.5 mm silica gel, left to air-dry and activated in an oven at 125 \textsuperscript{$\circ$}C for at least 4 h or overnight. The plates were then decontaminated by elution in DCM and activated at 125 \textsuperscript{$\circ$}C for 30 min. Maltenes (10 mg) were spotted on a plate with eicosane (C\textsubscript{20}H\textsubscript{42}; aliphatic), phenyldodecane (C\textsubscript{12}H\textsubscript{25}C\textsubscript{6}H\textsubscript{5}; monoaromatic) and anthracene (C\textsubscript{14}H\textsubscript{10}; triaromatic) elution standards and eluted in petroleum ether. The separated aliphatic and aromatic fractions were placed in short columns and eluted with petroleum ether and DCM respectively. The final fractions were reduced to 1 ml and analysed by GC-FID and GC-MS with heptadecylcyclohexane (C\textsubscript{23}H\textsubscript{46}) and \textit{p}-terphenyl (C\textsubscript{6}H\textsubscript{5}C\textsubscript{6}H\textsubscript{4}C\textsubscript{6}H\textsubscript{5}) as internal aliphatic and aromatic standards respectively.
\subsection*{Analytical instruments}
Maltene and RICO products were analysed using an Agilent 6890 instrument for gas-chromatography-flame ionisation detection (GC-FID) and an Agilent 7975C instrument for gas chromatography-mass spectrometry (GC-MS).\\
Iinitial compound screening was performed using the GC-FID equipped with a 30 m HP5-MS column (0.25 mm internal diameter, 0.25 $\mu$m polysiloxane stationary phase; J\&W Scientific, USA). Helium was used as carrier gas at a flow rate of 1 ml/min. The GC oven was initially held at 50 \textsuperscript{$\circ$}C for two minutes and then raised at a rate of 5 \textsuperscript{$\circ$}C/min to a final temperature of 310 \textsuperscript{$\circ$}C where it was held isothermally for 20 min. The instrument was run in splitless mode, whereby the injector was held at 280 \textsuperscript{$\circ$}C and the FID at 300 \textsuperscript{$\circ$}C.\\
Product detection was carried out using Agilent 7890A GC split/splitless injector at 280 \textsuperscript{$\circ$}C linked to an Agilent 7975C mass spectrometer. The oven was operated at the same temperature mode as GC-FID using a 30 m HP5-MS column (specification as above, J\&W Scientific, USA). A mass selective detector was used in selected ion monitoring and full scan modes (\textit{\textit{m/z}} 50–700). Compound identification was based on the NIST05 mass spectral library as well as comparison to mass spectra and relative retention times reported in other studies and in-house guides.
\subsection*{Ultrasonic velocity measurements}
Ultrasonic velocity measurements were performed at 25 $^{\circ}$C using the Resoscan Research System (TF Instruments). Resoscan is a high-resolution instrument allowing a simultaneous measurement of velocity and attenuation of liquid samples. The system includes a two-channel resonator unit with gold and lithium niobate piezocrystals, two 250 $\mu$l sample cells and a Peltier thermostat. Temperature control and measurement are performed via two external units enabling a resolution of 0.001 $^{\circ}$C and precision of $\pm$0.005 $^{\circ}$C. The small sample requirement of 170-250 $\mu$l and a high temperature stability enable to measure conformational changes on a molecular scale. In the vicinity of 25 $^{\circ}$C the sound velocity changes by around 5 ms\textsuperscript{-1}$^{\circ}$C\textsuperscript{-1}. Therefore, the precision limit of temperature corresponds to a systematic error in velocity of around 0.025 ms\textsuperscript{-1}, which is sufficiently small that it does not affect our results. The fundamental frequency of the instrument is 10 MHz, with range of 7-11.5 MHz, the precise frequency chosen to maximise the quality factor of the acoustic resonant cell.\\
Asphaltene solutions in toluene were injected using a glass syringe with an aluminium needle to avoid reactivity with the toluene solvent. The instrument was allowed to equilibrate for 3-4 min and left to take around 100 measurements for every solution concentration, from which an average velocity was determined. Aqueous surfactant solutions were injected an auto pipette. Up to 30 measurements were taken per concentration, with an equilibration time of 1 min.\\
During sample insertion an injector was held vertically parallel to the cell borehole and injection performed during 30-60 s to minimize air entrapment. The obtained data was analysed for extreme outliers, caused by air bubbles and other artefacts. Mean velocity values were plotted versus solute concentration to which piecewise linear regression models were fitted. 
\section*{Results}
\subsection*{Biodegradation of maltenes}
Biodegradation of hydrocarbons (examples are shown in Table~\ref{biodegr}) results in the removal of saturated and aromatic compounds\cite{milner1977,peters2005b}. The level (LVL) of petroleum biodegradation (1-10) was estimated using the Wenger \textit{et al.}~\cite{wenger2002,peters2005b} scales and diagnostic mass spectrometric assignments~\cite{peters2005a}. Gas chromatograms used in biodegradation analysis are provided in the Supplementary Information (Fig. S1-S3 online).\\
E1 has little signs of biodegradation (Fig. S1 (a-d) online). The \textit{m/z} 85 mass chromatogram shows a homologous series of \textit{n}-alkanes from which compounds ($<$C\textsubscript{11}) are absent. Regular isoprenoids, such as 2,6,10,14-tetramethylpentadecane (pristane) and 2,6,10,14-tetramethylhexadecane (phytane), are abundant (\textit{m/z} 183) as are hopanes and steranes (\textit{m/z} 191 and 218). Therefore, E1 is classified as LVL 2 biodegraded. The biodegradation extent of E2 (Fig. S1 (e-h) online) is greater than that of E1, indicated by the removal of \textit{n}-decane from the \textit{n}-alkanes. The isoprenoid, hopane and sterane distributions (\textit{m/z} 183, 191 and 218) are very similar to E1. Therefore, E2 is is classed as LVL 2-3 biodegraded. In contrast, E3 illustrates a complete removal of \textit{n}-alkanes which puts it into the heavily biodegraded range (Fig. S2 (a-f) online). Hopanes (\textit{m/z} 191) are partially removed and 25-norhopanes have been formed. Sterane compounds (\textit{m/z} 218) are difficult to resolve and benzoalkanes are degraded (\textit{m/z} 91). However, monoaromatic steroids (\textit{m/z} 253) are present, therefore E3 is estimated to be LVL 6-7 biodegraded. E4 shows signs of early biodegradation (Fig. S2 (g,h) and S3 online) with the homologous \textit{n}-alkane series starting from \textit{n}-C\textsubscript{11}. Isoprenoids, such as pristane and phytane, are detectable. Hopanes and steranes are abundant and only C\textsubscript{29}-norhopane has formed. Therefore, E4 is estimated to be LVL 3 biodegraded.
\subsection*{RICO of asphaltenes}
The products analysed from RICO are homologous series of \textit{n}-alkanoic and $\alpha,\omega$-\textit{di}-\textit{n}-alkanoic fatty acid \textit{bis}-methyl esters (FAME, \textit{m/z} 74; DFAME, \textit{m/z} 98)~\cite{peng1999}. Partial and total ion chromatograms (TICs) are provided in Supplementary Information. Table~\ref{rico} includes abundances of FAME and DFAME compounds in RICO products.\\
In all cases the TICs are dominated by FAME series. The TICs for E1 and E2 (Fig. S4 online) illustrate left-skewed distributions with a mild even-odd predominance of medium molecular weight compounds. We calculated the ratio of even-odd medium molecular-weight \textit{n}-alkanoic acids (C\textsubscript{12-18} to C\textsubscript{11-17}) as an indicator of biological activity~\cite{tissot1984} (Table~\ref{rico}), which was 1.711 and 1.33 for E1 and E2 respectively. Note that the high C\textsubscript{16} and C\textsubscript{18} peaks also drive the even-odd predominance of E1 which is highest out of the four samples. The DFAME compounds at C\textsubscript{4}-C\textsubscript{6} are mixed with the baseline and become increasingly more resolved after C\textsubscript{7}. In comparison, the compound distributions of E3 and E4 (Fig. S4-S5 online) are more peaked with varying compound resolution. Even-odd ratios are 1.349 and 1.446 respectively. The DFAME series are very well-resolved for E3 but are quite weak for E4. From Table~\ref{rico} it is evident that the FAME and DFAME compounds are negatively correlated. As nanoaggregation kinetics of asphaltenes are affected by steric hindrance between side-chains~\cite{buenrosto2001,mullins2010,mullins2011} we will test the coupling between asphaltene FAME distributions and nanoaggregation results observed in ultrasonic velocity measurements in the Discussion section.
\subsection*{Resoscan ultrasonic velocity measurements}
\begin{figure}[ht!]
\centering
\includegraphics[scale=0.6,angle=270]{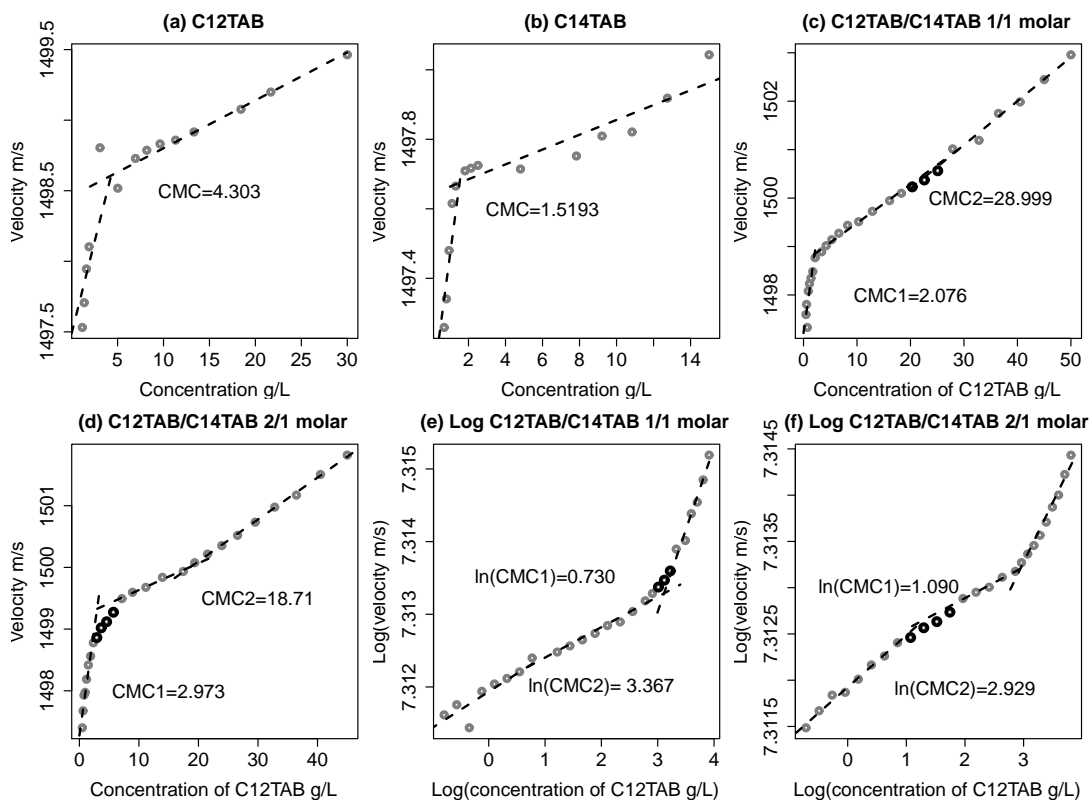}
\caption{Concentration-velocity measurements of CTAB pure and mixed aqueous solutions. Dashed lines represent fitted linear regressions, $R^2$ values are reported in Table~\ref{ctab}. In plots c-f, CMC1 and CMC2 refer to primary and secondary micelle formation respectively. Points marked in black indicate data that were not included in linear regression estimation.} 
\label{ctab1}
\end{figure}

Aqueous solutions of C\textsubscript{12}TAB, C\textsubscript{14}TAB and their mixtures (C\textsubscript{12}TAB$/$C\textsubscript{14}TAB, 1/1 and 2/1 molar) were used to verify the sensitivity of the ultrasonic velocity instrument to micellarisation. Figure~\ref{ctab1} illustrates velocity-concentration plots with superimposed linear models whose $R^2$ values are provided in Table~\ref{ctab}. Every concentration measurement in an average of ca. 10 points. Confidence intervals are not shown as the standard deviations are below image resolution, their mean values are shown in Table~\ref{ctab}. Plots a,b and the corresponding $R^2$ values provide good evidence of the linearity between velocity and surfactant concentration, and the CMC values correspond to previous measurements~\cite{watson2004,sun2005,zielinski1986}. Plots c,d illustrate multiple micellarisation in CTAB mixtures, with a strong indication of the primary critical micelle concentration (CMC\textsubscript{1}); the secondary critical micelle concentration (CMC\textsubscript{2}) may be highlighted by taking a (natural) log-transformation as shown in plots e,f. Similar results were found by Ray \textit{et al.}~\cite{ray2005} illustrating multiple micellarisation of CTAB surfactants using tensiometric, conductometric and other methods.\\
\begin{figure}[!ht]
\centering
\includegraphics[scale=0.6]{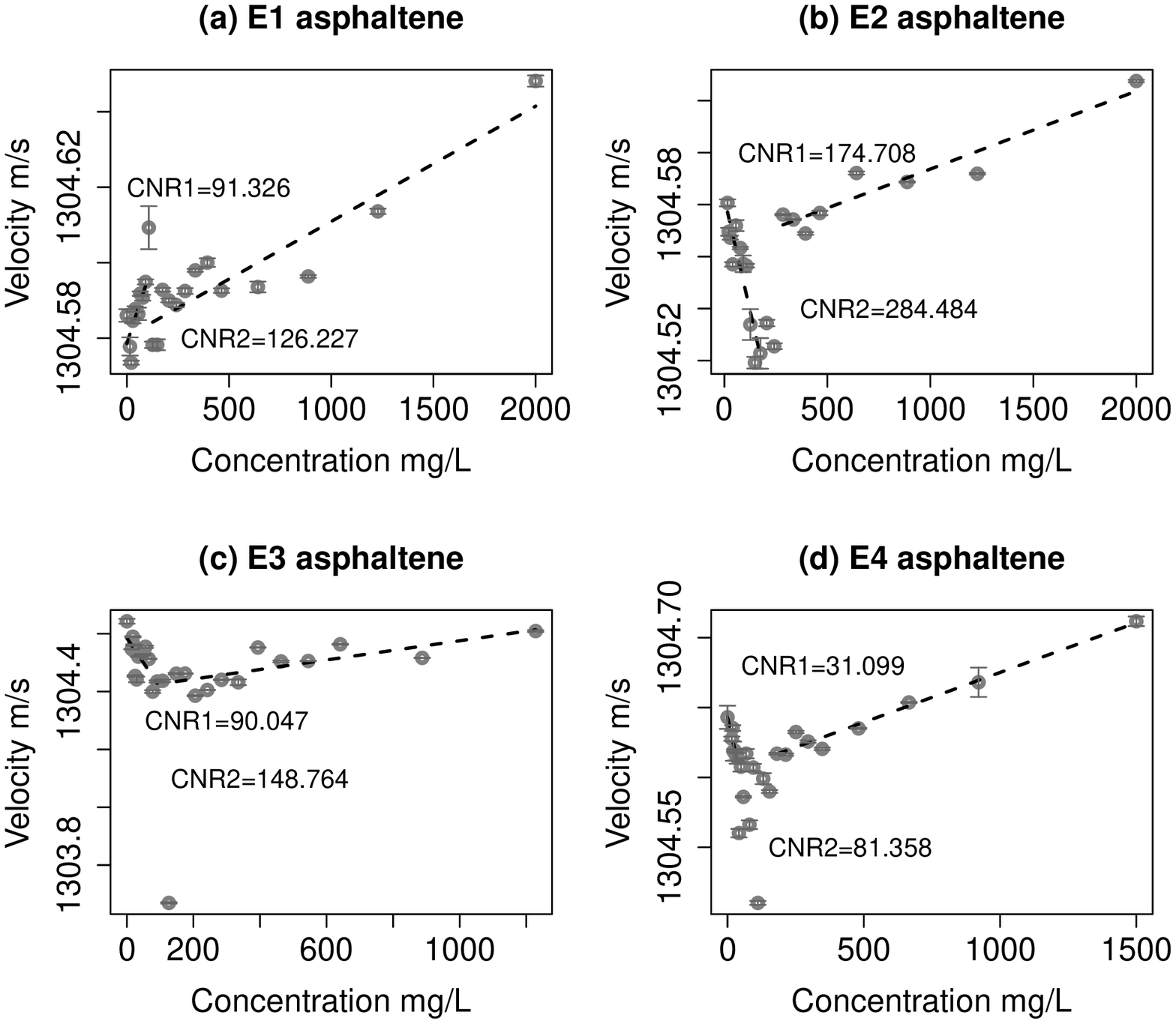}
\caption{Concentration-velocity measurements of asphaltene-toluene mixtures. Dashed lines illustrate estimated linear models using constrained optimisation, CNR1 and CNR2 refer to the onset and decline of the critical nanoaggregation region respectively.}
\label{asphaltene}
\end{figure}
Figure~\ref{asphaltene} illustrates ultrasonic velocity measurements of four asphaltene samples with linear models superimposed. Every concentration measurement is an average of 60-100 points, the confidence intervals represent the mean $\pm$ standard deviation. Piecewise regression models indicated by dashed lines were fitted using a constrained optimisation scheme~\cite{bertsekas1982,smith1997} which is detailed in the Supplementary Information. When choosing between different regression estimations we deploy this scheme as a guide to indicate where linear behaviour becomes no longer plausible by excessively penalising the $R^2$ measure given large outliers and changes in regression slope. This should be taken into account when interpreting the penalised $R^2$ values provided in Table~\ref{asph}, which are expected to be low. In general, the four plots suggest that at either ends of asphaltene concentration range the association with ultrasonic velocity is linear, and the CNR indicated by large outliers. The width of the CNR ($\Delta$ CNR) and the velocity difference between the monomeric and aggregated linear models ($\Delta \, v$) are is highly-variable across the samples and are discussed in the next section. Note that the change in the speed of sound is larger than the 95\% confidence intervals in the majority of cases. The larger error bars are likely to occur due to trace amounts of dissolved air.
\section*{Discussion}
\begin{figure}[ht!]
\centering
\includegraphics[scale=0.6,angle=270]{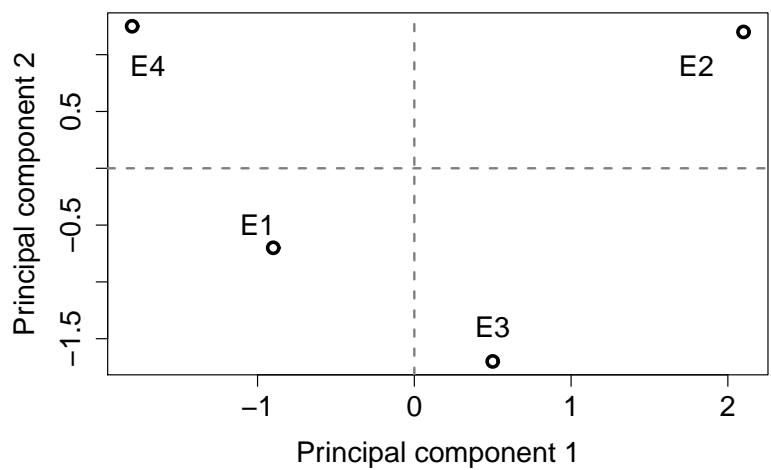}
\caption{Division of asphaltene samples based on PC1 and PC2.}
\label{score}
\end{figure}

The sample E1 is estimated to have the narrowest $\Delta$ CNR of 34.901 mg/L and is also accompanied by the smallest $\Delta\, v$ of 0.017 m/s. In contrast, E2 has the largest $\Delta$ CNR of 106.776 mg/L and $\Delta\, v$ of 0.053 m/s. The total $R^2_p$ values for two samples are similar. The sample E3 has the lowest total $R^2_p$ of 0.4435 although its $\Delta$ CNR and $\Delta\, v$ are 58.717 and 0.027 m/s which is not the most extreme. E4 has the highest aggregated $R^2$ and the earliest onset of nanoaggregation at 31.099 mg/L, although $\Delta$ CNR is estimated to be twice as great as the monomeric region. The trend below and above the CNR is significant over the measurement error bars for all samples. These trends are similar to those observed in ultrasonic characterisation of Tween 80-toluene mixtures and asphaltene-toluene mixtures by Andreatta, \textit{et al.}~\cite{andreatta2005}. In particular, the gradient change is consistent across our four samples, except E1 where both slopes are positive. We are not sure what the reason is for this discrepancy, but a few things can be noted here. The theoretical linear model underlying our measurements that relates the speed of sound to asphaltene concentration~\cite{zielinski1986,andreatta2005} is a function of apparent molar densities and compressibilities. As asphaltenes constitute a petroleum fraction with a range of molecular properties, isolating apparent molar quantities is challenging if not impossible. Understanding the chemico-physical cause of changing gradient in asphaltene ultrasonic velocity measurement would be a very interesting task, however it goes beyond the scope of present investigation.\\
We performed principal component analysis (PCA)~\cite{mardia1979,cox2005} to assess the combined impact of the above variables on asphaltene nanoaggregation. The variables used in PCA are $\Delta$ CNR, $\Delta\, v$, $R^2_p$ and percentage of FAME C\textsubscript{11-18} and FAME C\textsubscript{$\geq$19} out of fatty acid methyl ester products. It is uncertain whether the longer DFAME compounds are PAH linkages as there is no evidence in the literature that inter-aromatic bridges can be over seven carbons long~\cite{schuler2015,schuler2017}. Therefore, we will exclude this information from PCA. Lower-molecular weight FAMEs (C$_{<7}$) are highly-volatile and can be partially/completely lost during the RICO procedure~\cite{ma2008,peng1999}, thus will be excluded from the following inference as well. Although the latter is unfortunate, we assume that their contribution is relatively insignificant to hindrance effects as compared to the longer appendages. Table~\ref{loadings} shows the loadings of the principal components (PCs), Minitab software was used for this analysis. Figure~\ref{score} illustrates sample division based on PC1 and PC2.
Here, principal component 1 (PC1) gives the largest weighting to $\Delta$ CNR, $\Delta\, v$ and the long aliphatic chains assigning the lowest absolute weight to $R^2_p$. This component can, therefore, be interpreted as the aggregation complexity variable, drawing a positive relation between the size of the CNR and the amount of longer asphaltene side-chains, all of which negatively affect the linear model fit. Observing the distribution in Figure~\ref{score}, PC1 indicates that aggregation complexity is lowest for E4 and highest for E2 which is consistent with the aggregation behaviour in Figure~\ref{asphaltene}. The principal component 2 (PC2) gives the largest weighting to $R^2_p$ whilst making a polar contrast to medium-length asphaltene side-chains. Interestingly, the contrast to long side-chains is half that of medium-sized ones, which implies that the linear model fit is more affected by the abundance of medium-length side-chains or perhaps the size variability. PC2 can be interpreted as the ‘optimal statistical performance’ component. In this domain, E4 and E2 have a very similar performance which is consistent with their fits to linear models (Figure~\ref{asphaltene}). Together, PC1 and PC2 explain 98\% of variation in the data suggesting a link between asphaltene structural and aggregation data.
\section*{Conclusions}
In conclusion, we have provided evidence for an asphaltene CNR. Asphaltene samples were selected to represent a range of molecular architectures and biodegradation levels of the parent oils. Their geochemical properties were obtained using maltene analysis, RICO and GC-MS. Ultrasound velocity characterisation was used to detect nanoaggregation in conjunction with the model by Zielinski, \textit{et al.}~\cite{zielinski1986}. The analysis of asphaltene solutions in toluene illustrated a variability of aggregation behaviours, subject to asphaltene structural properties. This relationship was confirmed by PCA of asphaltene coupled structural and velocity data. We found that asphaltene structural properties, in particular longer aliphatic appendages and increased variation in size, contribute to a wider aggregation region and a decreased fit to the two-regression model. The results call for a further investigation whereby the origin of longer DFAME compounds is established and their contribution to nanoaggregation also resolved. Ultrasonic velocity results will also be used to construct a probabilistic model of the CNR.
\bibliography{ASbibliography}
\section*{Acknowledgments}
We thank Dr Melvin Holmes, laboratory technicians David Earley, Bernard Bowler and Dr Paul Donohoe for their kind help, and Dr Jian Qing Shi for assistance in statistical inference. This work is funded by the Natural Environment Research Council and Newcastle University (1:1) and is part of the Centre for Doctoral Training in Oil and Gas.
\section*{Author contributions statement}
G.D.A. and M.J.W.P. conceived the project. N.G.P., M.J.W.P. and G.D.A. conceived the ultrasonic characterisation experiments. N.P. provided expertise for ultrasonic characterisation experiments. G.D.A. and A.S. conceived the RICO and biodegradation experiments. G.D.A. provided expertise for GC-MS analysis. A.S. performed the experiments, analysed the results and wrote the manuscript. N.G.P and G.D.A. made the main contribution to the critique of the manuscript. All authors reviewed the manuscript.
\section*{Additional information}
\subsection*{Competing financial interests}
The authors declare no competing financial interests.
\subsection*{Data availability statement}
The datasets generated during and/or analysed during the current study are available from the corresponding author on reasonable request.
\begin{table}[!ht]
\centering
\begin{tabular}{|l|l|}
\hline
Number& Name\\ \hline
1 &\textit{n}-undecane\\ \hline
2 & 2,6,10,14-tetramethylpentadecane (pristane)\\ \hline
3 & 2,6,10,14-tetramethylxehadecane (phytane)\\ \hline
4 & heptadecylcyclohexane\\ \hline
5~\cite{brincat2001} & (C\textsubscript{19}) tricyclic terpane\\ \hline
6~\cite{brincat2001} & (C\textsubscript{21}) tricyclic terpane\\ \hline
7~\cite{brincat2001} & (C\textsubscript{23}) tricyclic terpane\\ \hline
8~\cite{brincat2001} & (C\textsubscript{25}) tricyclic terpane\\ \hline
9~\cite{brincat2001} & (C\textsubscript{27}) 17$\alpha$-22,29,30-\textit{trinor}-hopane\\ \hline
10~\cite{brincat2001} & (C\textsubscript{29}) $17\alpha,21\beta$-\textit{nor}-hopane\\ \hline
11~\cite{brincat2001} & (C\textsubscript{30}) $17\alpha,21\beta$-\textit{nor}-hopane\\ \hline
12~\cite{brincat2001} & (C\textsubscript{31}) $17\alpha,21\beta$-\textit{homo}-hopane 22\textit{S} and 22\textit{R} epimers\\ \hline
13~\cite{brincat2001} & (C\textsubscript{35}) $17\alpha,21\beta$-\textit{pentakishomo}-hopane 22\textit{S} and 22\textit{R} epimers\\ \hline
14~\cite{brincat2001} & (C\textsubscript{27}) $5\alpha,14\beta,17\beta$-cholestane 20\textit{R} and 20\textit{S} epimers\\ \hline
15~\cite{brincat2001} & (C\textsubscript{28}) $5\alpha,14\beta,17\beta$-24-methylcholestane 20\textit{R} and 20\textit{S} epimers\\ \hline
16~\cite{brincat2001} & (C\textsubscript{29}) $5\alpha,14\beta,17\beta$-24-ethylcholestane 20\textit{R} and 20\textit{S} epimers\\ \hline
17 & 1,4-diphenylbenzene (\textit{p}-terphenyl)\\ \hline
18~\cite{abbott1985,abbott1985a} & (C\textsubscript{27}$+$C\textsubscript{28}) C-ring monoaromatic steroid\\ \hline
19~\cite{abbott1985,abbott1985a} & (C\textsubscript{27}$+$C\textsubscript{28}$+$C\textsubscript{29}) C-ring monoaromatic steroid\\ \hline
20~\cite{abbott1985,abbott1985a} & (C\textsubscript{29}) C-ring monoaromatic steroid\\ \hline
\end{tabular}
\caption{Compound table for biodegradation of maltenes and Supplement Fig. S1-S3.}
\label{biodegr}
\end{table}
\begin{table}[!ht]
\centering
\begin{tabular}{|l|l|l|l|l|l|}
\hline
Sample Name & \% FAME & \% FAME C\textsubscript{11-18} & \% FAME C\textsubscript{$\geq$19} & FAME $\text{C}_{12-18}/\text{C}_{11-17}$ & \% DFAME \\ \hline
E1 & 69.5 & 62.5784 & 13.7502 & 1.711 & 9.6\\ \hline
E2 & 71.0 & 58.9249 & 20.8198 & 1.33 & 2.4 \\ \hline
E3 & 66.7 & 65.4490 & 17.6684 & 1.349 & 12.9\\ \hline
E4 & 71.8 & 54.4096 & 10.6304 & 1.446 & 4.2\\ \hline
\end{tabular}
\caption{Abundance of FAME and DFAME compounds in RICO products. Entries \% FAME C\textsubscript{11-18} and \% FAME C\textsubscript{$\geq$19} refers to percentage of medium- and long-chain compounds out of total FAME products.}
\label{rico}
\end{table}
\begin{table}[!ht]
\centering
\begin{tabular}{|l|l|l|l|l|l|l|}
\hline
Sample name & SD & CMC\textsubscript{1} & CMC\textsubscript{2} & $R^2_{\text{mono}}$ & $R^2_{\text{aggr}}$ & $R^2_{\text{inter}}$\\ \hline
C\textsubscript{12}TAB & 0.00037 & 4.303 g/L & NA & 0.6218 & 0.9699 & NA\\ \hline
C\textsubscript{14}TAB & 0.00075 & 1.519 g/L & NA & 0.8018 & 0.8379 & NA\\ \hline
C\textsubscript{12}TAB/C\textsubscript{14}TAB 1/1 molar & 0.00077 & 2.076 g/L & 28.999 g/L & 0.7386 & 0.9823 & 0.9839\\ \hline
C\textsubscript{12}TAB/C\textsubscript{14}TAB 2/1 molar & 0.00049 & 2.973 g/L & 18.710 g/L & 0.9365 & 0.9988 & 0.9844\\ \hline
\end{tabular}
\caption{Summary of CTAB concentration-velocity data. Mean sample standard deviation is denoted SD, subscripts of $R^2$ refer to models fitted in the estimated monomer (mono), aggregated (aggr) and CMC\textsubscript{1}-CMC\textsubscript{2} intermediate (inter) regions.}
\label{ctab}
\end{table}
\begin{table}[!ht]
\centering
\begin{tabular}{|l|l|l|l|l|l|l|l|}
\hline
Sample name & CNR\textsubscript{1} & CNR\textsubscript{2} & $\Delta$ CNR & $\Delta\, v$ & $R^2_{\text{mono}}$ & $R^2_{\text{aggr}}$ & Total $R^2_p$\\ \hline
E1 & 91.326 mg/L & 126.227 mg/L & 34.901 & 0.017 & 0.2252 & 0.5903 & 0.8155\\ \hline
E2 & 174.708 mg/L & 284.484 mg/L & 106.776 & 0.053 & 0.5329 & 0.5469 & 1.0798\\ \hline
E3 & 90.047 mg/L & 148.764 mg/L & 58.717 & 0.027 & 0.1064 & 0.3371 & 0.4435\\ \hline
E4 & 31.099 mg/L & 81.358 mg/L& 50.259 & 0.003 & 0.4295 & 0.6400 & 1.0695\\ \hline
\end{tabular}
\caption{Regression \textbf{penalised} $R^2$ values of asphaltene concentration-velocity data. Total $R^2_p$ denotes the sum of $R^2_{\text{mono}}$ and $R^2_{\text{aggr}}$, $\Delta$ CNR denotes the CNR width, $\Delta\, v$ denotes the velocity jump. Subscripts of CNR denote the onset\textsubscript{1} and decline\textsubscript{2} of aggregation. Penalised $R^2$ subscripts refer to estimated models in the monomer and aggregate regions.}
\label{asph}
\end{table}
\begin{table}[!ht]
\centering
\begin{tabular}{|l|l|l|l|l|l|}
\hline
Variable & PC1 & PC2 & PC3 & PC4 & PC5\\ \hline
$\Delta$ CNR & 0.516 & 0.322 & 0.523 & -0.009 & -0.597\\ \hline
$\Delta\,v$ & 0.587 & 0.060 & -0.336 & -0.688 & 0.256\\ \hline
$R^2_p$ & -0.032 & 0.681 & -0.644 & 0.261 & -0.228\\ \hline
FAME C\textsubscript{11-18} & 0.202 & -0.649 & -0.446 & 0.121 & -0.569\\ \hline
FAME C\textsubscript{$\geq$19} & 0.590 & -0.082 & -0.005 & 0.666 & 0.449\\ \hline
\end{tabular}
\caption{Loadings of the first five principal components.}
\label{loadings}
\end{table}
\end{document}